\documentclass[10pt,twocolumn,english,aps,preprint,showpacs]{revtex4}
\usepackage[T1]{fontenc}
\usepackage[latin9]{inputenc}
\usepackage{amsmath}
\usepackage{graphicx}
\usepackage{amssymb}

\makeatletter
\@ifundefined{textcolor}{}
{%
 \definecolor{BLACK}{gray}{0}
 \definecolor{WHITE}{gray}{1}
 \definecolor{RED}{rgb}{1,0,0}
 \definecolor{GREEN}{rgb}{0,1,0}
 \definecolor{BLUE}{rgb}{0,0,1}
 \definecolor{CYAN}{cmyk}{1,0,0,0}
 \definecolor{MAGENTA}{cmyk}{0,1,0,0}
 \definecolor{YELLOW}{cmyk}{0,0,1,0}
 }

\makeatother

\makeatother

\usepackage{babel}

\makeatother

\usepackage{babel}

\begin{document}

\preprint{This line only printed with preprint option}

\title{$^{75}$As NMR study of antiferromagnetic fluctuations in Ba(Fe$_{1-x}$Ru$_{x}$)$_{2}$As$_{2}$}

\author{Tusharkanti Dey, P. Khuntia, A.V. Mahajan}

\email{mahajan@phy.iitb.ac.in}

\affiliation{Department of Physics, IIT Bombay, Powai, Mumbai 400076, India }

\author{Shilpam Sharma, A. Bharathi}

\affiliation{Condensed Matter Physics Division, Materials Science Group, Indira
Gandhi Centre for Atomic Research (IGCAR), Kalpakkam 603102, India}
\begin{abstract}
Evolution of $^{75}$As NMR parameters with composition and temperature
was probed in the Ba(Fe$_{1-x}$Ru$_{x}$)$_{2}$As$_{2}$ system
where Fe is replaced by isovalent Ru. While the Ru-end member was
found to be a conventional Fermi liquid, the composition ($x=0.5$)
corresponding to the highest {\normalsize $T\mathrm{_{c}}$} ($20$\,K)
in this system shows an upturn in $^{75}$As $\frac{1}{T_{1}T}$ below
about 80 K evidencing the presence of antiferromagnetic (AFM) fluctuations.
These results are similar to those obtained in another system with
isovalent substitution BaFe$_{2}$(As$_{1-x}$P$_{x}$)$_{2}$ {[}Y.
Nakai, T. Iye, S. Kitagawa, K. Ishida, H. Ikeda, S. Kasahara, H. Shishido,
T. Shibauchi, Y. Matsuda, and T. Terashima, Phys. Rev. Lett. \textbf{105},
107003 (2010){]} and point to the possible role of AFM fluctuations
in driving superconductivity. 
\end{abstract}


\maketitle
\textbf{1. Introduction}

The recent discovery of superconductivity in an iron based material
\cite{Kamihara-JACS-130-2008} LaFeAs(O$_{1-x}$F$_{x}$) with a superconducting
transition temperature $T_{\mathrm{c}}=26$\,K and soon after, the
increase of $T_{\mathrm{c}}$ to $43$\,K with applied pressure \cite{Takahashi-Nature-453-2008}
has reignited the interest in superconductors. The main motivating
factors for this interest are (i) the presence of Fe, which is normally
not considered conducive to superconductivity and (ii) the superficial
similarity to the high-$T\mathrm{_{c}}$ cuprates, in terms of the
existence of FeAs layers.

The parent compound BaFe$_{2}$As$_{2}$ is a semimetal which crystallizes
in a ThCr$_{2}$Si$_{2}$-type structure ($I4/mmm$) and exhibits
a spin-density-wave (SDW) transition \cite{Rotter-PRB-78-2008} at
$\sim140$\,K. Superconductivity is induced following electron/hole
doping \cite{Sefat-PRL-101-2008,Rotter-PRL-101-2008}, by applying
pressure externally \cite{Alireza-JPCM-21-2009}, by replacing As
with isovalent P \cite{Jiang-JPCM-21-2009}, or by replacing Fe with
isovalent Ru \cite{Sharma-PRB-81-2010}.

Heterovalent substitutions manifestly give rise to a change in the
carrier concentration in addition to other effects. Therefore, it
appears interesting to investigate isovalent substitutions which might
help to narrow down the relevance of various factors to superconductivity.
Perhaps with this motivation, Ru substitution at the Fe site was attempted
\cite{Sharma-PRB-81-2010,Thaler-PRB-82-2010,Rullier-Albenque-PRB-81-2010}.
In the reported works relevant to single crystals, the authors were
able to substitute Ru in place of Fe, however, only up to a limit
of $x=0.37$ (in \cite{Thaler-PRB-82-2010}) and $x=0.5$ (in \cite{Rullier-Albenque-PRB-81-2010}).
Very recently, Eom \textit{et al}. \cite{Eom-arxiv-2011} prepared
single crystals upto $x=0.7$. On the other hand, in the work on polycrystalline
Ba(Fe$_{1-x}$Ru$_{x}$)$_{2}$As$_{2}$ by S. Sharma \textit{et al}.
\cite{Sharma-PRB-81-2010} complete substitution was achieved. In
Sr(Fe$_{1-x}$Ru$_{x}$)$_{2}$As$_{2}$ as well Schnelle \textit{et
al}. \cite{Schnelle-PRB-79-2009} achieved full substitution of Fe
with Ru in polycrystalline samples. Ru substitution suppresses the
long-range antiferromagnetic transition and superconductivity appears.
The role played by antiferromagnetic spin fluctuations has been suggested
to be important towards a pairing mechanism.

From magnetic susceptibility measurements, Nath and co-workers \cite{Nath-PRB-79-2009}
have estimated the density-of-states at the Fermi level $D(\epsilon_{F})$
of BaRu$_{2}$As$_{2}$ to be $2.1$\,states/(eV formula unit). Thaler
\textit{et al}. have done a detailed study on single crystalline Ba(Fe$_{1-x}$Ru$_{x}$)$_{2}$As$_{2}$
samples where they have shown the similarity of its phase diagram
with that of BaFe$_{2}$As$_{2}$ with pressure \cite{Thaler-PRB-82-2010}.
Brouet \textit{et al}. \cite{Brouet-PRL-105-2010} have studied a
Ba(Fe$_{0.65}$Ru$_{0.35}$)$_{2}$As$_{2}$ single crystal using
electrical transport and photoemission spectroscopy and concluded
that the electron and hole concentrations were equal to each other
but double of those in BaFe$_{2}$As$_{2}$. The increase in carrier
concentration has been attributed to a large change in the band structure
as compared to undoped BaFe$_{2}$As$_{2}$. From thermal conductivity
measurements on Ba(Fe$_{0.64}$Ru$_{0.36}$)$_{2}$As$_{2}$ single
crystal and comparing their data with other doping samples, Qiu \textit{et
al}. \cite{Qiu-arxiv-2011} have shown that nodal superconductivity
induced by isovalent doping (P at As site and Ru at Fe site) have
the same origin. From electrical transport measurements, Eom \textit{et
al}. \cite{Eom-arxiv-2011} suggest that whereas there is a deviation
from Fermi-liquid behavior for compositions around the maximum $T_{C}$,
overdoped compositions tend towards a Fermi-liquid behavior. However,
due to the difficulty in preparing single crystals of BaRu$_{2}$As$_{2}$
no comparison exists between the superconducting compositions and
the Ru end member.

Nuclear magnetic resonance (NMR) has been instrumental in the case
of cuprates in detecting antiferromagnetic fluctuations, pseudo-gap,
and other features. In the FeAs compounds as well, NMR has been extensively
used to obtain a deeper understanding of various physics issues \cite{Ishida-review}.
In fact, in the isovalent substituted system BaFe$_{2}$(As$_{1-x}$P$_{x}$)$_{2}$,
$^{31}$P NMR was found to show the prominence of antiferromagnetic
fluctuations above $T\mathrm{_{c}}$ in the superconducting compositions
\cite{Nakai-PRL-105-2010}. However, up to now, no local probe investigations
have been reported which might help to understand the variation of
properties in the Ba(Fe$_{1-x}$Ru$_{x}$)$_{2}$As$_{2}$ system.
In this paper, we report $^{75}$As NMR measurements on Ba(Fe$_{1-x}$Ru$_{x}$)$_{2}$As$_{2}$
($x=0,0.25,0.5,1$) samples. From our measurements, we show that the
Ru-end member behaves like a conventional Fermi liquid. In contrast,
for the $x=0.5$ sample ($T_{\mathrm{c}}^{(\mathrm{onset})}=20$\,K
), $^{75}$As $\frac{1}{T_{1}T}$ is constant at high-$T$ but increases
with decrease in $T$ below 80 K suggesting the emergence of antiferromagnetic
fluctuations. Also, no evidence of heavy Fermi liquid behaviour in
Ba(Fe$_{1-x}$Ru$_{x}$)$_{2}$As$_{2}$ is found from our NMR measurements.

\textbf{2. Experimental Details}

Polycrystalline samples of Ba(Fe$_{1-x}$Ru$_{x}$)$_{2}$As$_{2}$
($x=0,0.25,0.5,1$) were prepared at IGCAR as detailed in \cite{Sharma-PRB-81-2010}.
Basic characterisation of the samples was done by x-ray diffraction,
magnetization, and resistivity measurements. For NMR measurements,
we tried to align the powder samples by mixing with Stycast $1266$
epoxy and then curing overnight in an external magnetic field $H=93.954$\,kOe.
$^{75}$As NMR measurements were carried out at IIT Bombay using a
Tecmag pulse spectrometer in a magnetic field of $93.954$\,kOe using
a room-temperature bore Varian superconducting magnet. Variable temperature
measurements were performed using an Oxford continuous flow cryostat,
using liquid nitrogen in the temperature range $80-300$\,K and using
liquid helium in the temperature range $4-80$\,K. The $^{75}$As
has nuclear spin $I=\frac{3}{2}$ ($100\%$ natural abundance) and
gyromagnetic ratio $\frac{\gamma}{2\pi}=7.2919$\,MHz/T. Spectra
were obtained by Fourier transform of the spin echo resulting from
a $\frac{\pi}{2}-\tau-\frac{\pi}{2}$ pulse sequence. Spin-lattice
relaxation time ($T_{1}$) was obtained by fitting the time dependence
of spin-echo intensity $m(t)$ with the formula \begin{equation}
1-m(t)/m(\infty)=A\, exp(-t/T_{1})+B\, exp(-6t/T_{1})\label{eq:T1 recovery}\end{equation}

following a $\frac{\pi}{2}-t-(\frac{\pi}{2}-\frac{\pi}{2})$ sequence
where the $\frac{\pi}{2}$ pulse duration is $\sim4\mu s$.

\textbf{3. Results and discussions}

X-ray powder diffraction revealed the presence of small amount (several
percent) of FeAs as also unreacted Fe and Ru present in the samples.
The average Ru content is expected to be less than the nominal amount
as was already found by others \cite{Rullier-Albenque-PRB-81-2010,Thaler-PRB-82-2010}.
Another point to note is that the x-ray diffraction peaks for our
$x=0.5$ sample are broad compared to those of the end-members. This
is likely due to a distribution of Ru content for this sample. Similar
broadened peaks were also found in Sr(Fe$_{1-x}$Ru$_{x}$)$_{2}$As$_{2}$
\cite{Qi-Physica C-469-2009}. The inability of Thaler \textit{et
al}. \cite{Thaler-PRB-82-2010} and Rullier-Albenque \textit{et al}.
\cite{Rullier-Albenque-PRB-81-2010} to obtain homogeneous single
crystals of Ba(Fe$_{1-x}$Ru$_{x}$)$_{2}$As$_{2}$ with $x>0.21$
is perhaps related to the difficulty in obtaining a sharply defined
composition. In any case, whereas a small amount of extrinsic impurity
or a distribution of Ru content does affect bulk properties like magnetisation
and resistivity, using a local probe such as NMR we will be able to
probe the instrinsic properties. %
\begin{figure}[h]
 \includegraphics[scale=0.4]{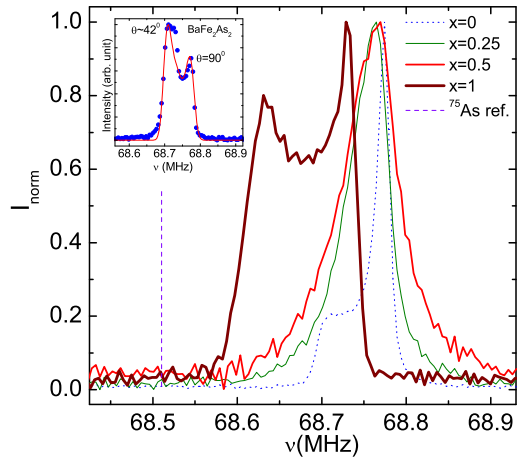}\caption{$^{75}$As NMR spectra measured at room temperature in a fixed field
$93.954$\,kOe for all four Ba(Fe$_{1-x}$Ru$_{x}$)$_{2}$As$_{2}$
samples. The $^{75}$As reference frequency is marked by a dashed
vertical line. Inset: $^{75}$As NMR spectrum for a randomly oriented
powder sample of BaFe$_{2}$As$_{2}$ (blue dots). The simulated spectrum
(red line) for the central line for $\nu_{Q}=3.028$\,MHz and $\eta=0$
(see text).}

\label{fig:BaFeRuAs2 Spectra}\centering{} 
\end{figure}

Resistivity and magnetic susceptibility (not shown here) evidence
the occurence of superconductivity in the BaFeRuAs$_{2}$ ($x=0.5$)
sample with a transition onset at $\sim20$\,K similar to \cite{Sharma-PRB-81-2010}.
Following this basic characterisation, we investigate the normal state
properties using $^{75}$As NMR as a local probe. Before describing
our measurements, we will first state some basic facts pertaining
to the NMR of $^{75}$As nuclei. In the Ba(Fe$_{1-x}$Ru$_{x}$)$_{2}$As$_{2}$
system, $^{75}$As ($I=\frac{3}{2}$) is not at a site of cubic symmetry.
This gives rise to a non-zero electric field gradient (EFG) at the
$^{75}$As site and coupled with its electric quadrupole moment, there
arise changes in the NMR lineshape as also the spin-lattice relaxation
behaviour. When the quadrupole term in the Hamiltonian is weak compared
to the Zeeman term it is enough to consider the effects upto first
order in perturbation theory. In this case, the central line ($\mbox{\ensuremath{\frac{1}{2}\leftrightarrow-\frac{1}{2}}}$
transition) is unaffected while satellite lines appear corresponding
to the $\mbox{\ensuremath{-\frac{3}{2}\leftrightarrow-\frac{1}{2}}}$
and $\mbox{\ensuremath{\frac{3}{2}\leftrightarrow\frac{1}{2}}}$ transitions.
The positions of the satellite lines depends on the angle $\theta$
between the magnetic field direction and the direction made by the
maximum of the EFG $V_{zz}$. When quadrupole effects are considered
to second order (and for axial symmtery), the central line position
(in the absence of anisotropy) also depends on $\theta$ and is given
by the following equation

\begin{equation}
\nu_{(\pm\frac{1}{2})}^{(2)}=\nu_{0}+\frac{\nu{}_{Q}^{2}}{32\nu_{0}}\left[I(I+1)-\frac{3}{4}\right](1-cos^{2}\theta)(9cos^{2}\theta-1)\label{eq:lineshape}\end{equation}

where $\nu_{Q}$ is the quadrupole frequency and $\nu_{o}$ the Larmor
frequency. In a randomly aligned polycrystalline sample the central
lineshape is the powder average, resulting in two peaks corresponding
to $\theta\thickapprox41.8{}^{o}$ and $\theta=90^{o}$. It is known
\cite{Baek-PRB-78-2008} that $\nu_{Q}$ for BaFe$_{2}$As$_{2}$
is about $3$\,MHz. The central line pattern in our randomly aligned
BaFe$_{2}$As$_{2}$ sample can be generated (see inset of figure
\ref{fig:BaFeRuAs2 Spectra}) taking $\nu_{Q}=3.028$\,MHz. In going
from BaFe$_{2}$As$_{2}$ to BaRu$_{2}$As$_{2}$ these numbers are
expected to change only nominally.

We have attempted to align the samples with the $ab$-plane in the
applied field direction. As seen from the contrast between the figure
\ref{fig:BaFeRuAs2 Spectra} inset (randomly aligned) and the curve
for $x=0$ (aligned sample) in the main figure \ref{fig:BaFeRuAs2 Spectra},
this is well achieved for BaFe$_{2}$As$_{2}$. However, it appears
that the Ru-substituted samples did not align in the field. This arises
perhaps due to the absence of single crystallites in the powders.
Nevertheless, our objectives from the spectra measurements are twofold;
(i) to determine the Knight shift as a function of temperature and
(ii) to irradiate the central line and then determine the spin-lattice
relaxation rate. From the position of the $\theta=90^{o}$ peak, the
shift can be determined. Since the central line is only about $200$\,kHz
in overall extent whereas our typical$\frac{\pi}{2}$ pulsewidth is
$4\mu s$, there is sufficient spectral width to saturate the central
line with a single $\frac{\pi}{2}$ pulse.

\begin{figure}
\includegraphics[scale=0.4]{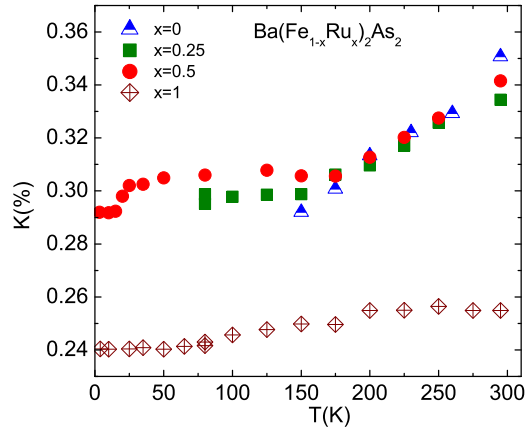}\caption{Shift for various compositions is shown as a function of temperature
$T$. For the metallic one ($x=1$) the shift $K$ is nearly independent
of $T$. For $x=0.5$ the superconducting transition is manifested
by a step like change in the $K$. }

\label{fig:BaFeRuAs2 Shift}\centering{} 
\end{figure}

The measured spectra could be simulated assuming $K_{x}=K_{y}\thickapprox K_{z}=K$,
$\nu_{Q}=3.028$\,MHz and $\eta=0$ (i.e., $V_{xx}=V_{yy}$). The
isotropic shift is therefore determined using equation \ref{eq:lineshape}
($K=\frac{\nu_{0}-\nu_{ref}}{\nu_{ref}}$). $K$ gives useful information
about the intrinsic susceptibility (having both spin and orbital contributions)
of the sample. We have $K=K_{spin}+K_{chem}$, where $K_{spin}=\frac{A_{hf}\chi_{spin}}{N_{A}\mu_{B}}$
is the spin part of the Knight shift and $K_{chem}$ is the temperature
independent chemical shift. The hyperfine coupling is $A_{hf}$ while
$N_{A}$ and $\mu_{B}$ are the Avogadro number and the Bohr magneton,
respectively. The temperature variation of shift is shown in figure
\ref{fig:BaFeRuAs2 Shift}. As to the value of the chemical shift,
there is some disagreement in literature. Ning \textit{et al}. \cite{Ning-JPSJ-78-2009}
estimated the chemical shift from their measurement at $4.2$\,K.
On the other hand, Kitagawa \textit{et al}. \cite{Kitagawa-JPSJ-77-2008}
obtained chemical shift by plotting $K-\chi$ and then extrapolating
to zero susceptibility. Kitagawa \textit{et al}. commented that the
value of $K_{chem}\sim0.2\%$ (also found by Ning \textit{et al}.)
seemed rather large and perhaps the extrapolation was not reliable.
In view of this as also the possibility that $K_{chem}$ might be
composition dependent, we have not corrected our shift data for the
chemical shift. %

For $x=1$, the shift is almost independent of temperature like a
conventional Fermi liquid. For the other three samples (including
the superconducting composition) the shift value is higher than that
for $x=1$. Further it increases linearly with temperature above $160$\,K.
Below $160$\,K the shift is independent of temperature for $x=0.25$
and $x=0.5$ samples. The drop in shift at $\sim$$20$\,K for $x=0.5$
is due to the superconducting transition. The SDW transition is seen
at $140$\,K and $70$\,K for $x=0$ and $0.25$, respectively.
The nearly unchanged shift in going from $x=0$ to $x=0.5$ indicates
that the density-of-states at the Fermi level $D(\epsilon_{F})$ remains
unchanged with substitution up to $x=0.5$. On the other hand, the
non-superconducting, metallic end-member has a significantly smaller
shift (and therefore $D(\epsilon_{F})$). In case $K_{chem}$ is taken
to be about $0.2\%$ (as in \cite{Ning-JPSJ-78-2009}) the reduction
in $K_{spin}$ is by a factor of two in going from $x=0.5$ to $x=1$.
A similar trend is seen in BaFe$_{2}$(As$_{1-x}$P$_{x}$)$_{2}$
for the larger values of $x$ beyond the superconducting dome \cite{Nakai-PRL-105-2010}.

\begin{figure}
\includegraphics[scale=0.45]{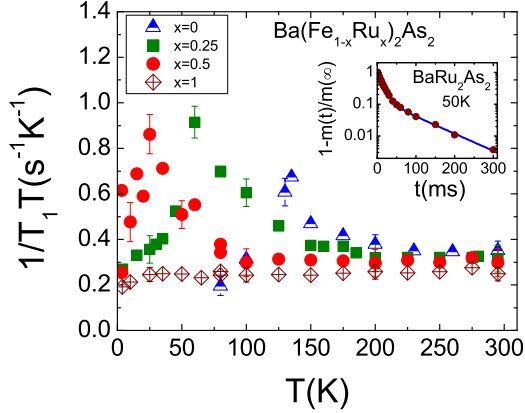}\caption{$^{75}$As spin-lattice relaxation rate divided by temperature ($\frac{1}{T_{1}T}$)
is shown as a function of temperature (typical error bars are shown).
Inset: The recovery of the $^{75}$As nuclear magnetization (wine
color dots) for BaRu$_{2}$As$_{2}$ at $50$\,K and its fit (blue
line) with Eq. \protect\ref{eq:T1 recovery} with $A=0.1$ and $B=0.9$. }

\label{fig:BaFeRuAs2 T1inv}\centering{} 
\end{figure}

To study the low-energy spin dynamics, $T_{1}$ is measured as a function
of temperature for all the samples by the saturation recovery method.
A representative data-set for BaRu$_{2}$As$_{2}$ at $50$\,K and
its fit with equation \ref{eq:T1 recovery} is shown in the inset
of figure \ref{fig:BaFeRuAs2 T1inv}. %

\begin{figure}
\includegraphics[scale=0.3]{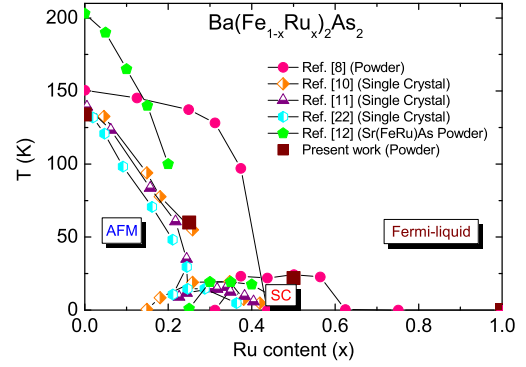}\caption{\label{fig:PhaseDiagram} Phase diagrams for powder (solid symbol)
and single crystal (half-filled symbol) samples of Ba(Fe$_{1-x}$Ru$_{x}$)$_{2}$As$_{2}$
available in literature, are shown. Phase diagram Sr(Fe$_{1-x}$Ru$_{x}$)$_{2}$As$_{2}$
powder sample is also shown. Different regions of the phase diagram
are marked based on the NMR parameters obtained from our measurements.}

\end{figure}

For the $x=1$ sample, the data could be fit with $A:B=1:9$ in the
full temperature range as also for other compositions for temperatures
higher than about $60$\,K. At lower temperatures the ratio of the
coefficients is about $4:6$. The ratio of the coefficients is expected
to be different from $1:9$ in case the relaxation is not magnetic.
In the present case, it is possible that the lattice becomes soft
prior to the superconducting transition and then quadrupolar relaxation
also contributes. However, there could be other possibilities. The
variation of the relaxation rate divided by temperature as a function
of temperature is shown in figure \ref{fig:BaFeRuAs2 T1inv}. Note
that conventional, wide-band metallic systems will have a $T$-independent
$K_{spin}$. Therefore, Korringa behaviour ($T$-independent $K{}_{spin}^{2}T_{1}T$)
in a conventional metal implies the constancy of $\frac{1}{T_{1}T}$
with $T$. For the non-superconducting $x=1$ composition, $\frac{1}{T_{1}T}$
is seen to be independent of $T$ down to 4 K. It's shift is also
found to be only weakly $T$-dependent. Of course, one needs to know
$K_{chem}$ in order to make a quantitative comment about the $T$-variation
of $K_{spin}$ and consequently of ${K{}_{spin}^{2}T_{1}T}$. Since
the present problem involves transferred hyperfine interactions, the
relaxation rate may also be affected by the form factor ($q$-dependence
of hyperfine couplings) which could lead to a deviation from the free-electron
value of ${K{}_{spin}^{2}T_{1}T}$ as also its expected $T$-independence.
Our observed value of ${K{}_{spin}^{2}T_{1}T}$ for $x=1$ is within
an order of magnitude of the free electron value and is also nearly
$T$-independent which, keeping the above limitations in mind, seems
to suggest the validity of the Fermi-liquid picture for BaRu$_{2}$As$_{2}$.
It's worth mentioning that Eom \textit{et al}. \cite{Eom-arxiv-2011}
also found Fermi-liquid behavior for higher doping ($x\sim0.7$) samples.
In contrast, for the superconducting sample ($x=0.5$) ($\frac{1}{T_{1}T}$)
is independent of $T$ down to about 80 K below which it shows an
upturn before dropping at the superconducting transition at $T\sim20$
K. The upturn in ($\frac{1}{T_{1}T}$) as a function of temperature
has also been seen (in the context of iron pnictides) in BaFe$_{2}$(As$_{1-x}$P$_{x}$)$_{2}$
in $^{31}$P NMR studies \cite{Nakai-PRL-105-2010} and is believed
to signify the existence of antiferromagnetic fluctuations. For $x=0$
and 0.25, ($\frac{1}{T_{1}T}$) shows an anomaly at the SDW transition
which is also seen in shift measurements. Compiling the published
data on Ba(Fe$_{1-x}$Ru$_{x}$)$_{2}$As$_{2}$ (single crystals
and powders), we summarize its magnetic phase diagram in figure \ref{fig:PhaseDiagram}.
Data for Sr(Fe$_{1-x}$Ru$_{x}$)$_{2}$As$_{2}$ are also shown for
reference. The heavily overdoped region exhibits Fermi-liquid behavior
and near the top of the superconducting dome, there is a prominence
of AFM fluctuations along with non-Fermi-liquid behavior based on
the NMR data. The AFM/SDW part shows anomalies in the shift at the
ordering temperature as expected.

\textbf{4. Conclusions}

We have reported $^{75}$As NMR measurements on Ba(Fe$_{1-x}$Ru$_{x}$)$_{2}$As$_{2}$
($x=0,0.25,0.5,1$) samples for the first time. The full ruthenated
sample ($x=1$) is a conventional Fermi liquid as evidenced from shift
and $T_{1}$ measurements. On moving towards the superconducting composition
($x=0.5$), the density-of-states at the Fermi level increases. Further,
for the $x=0.5$ sample, $\frac{1}{T_{1}T}$ shows an upturn with
decreasing temperature indicative of the emergence of antiferromagnetic
fluctuations. Our results suggest the importance of AFM fluctuations
in the superconducting mechanism in the FeAs -based systems. Similar
results have been reported for isovalent P-substitution at the As
site \cite{Nakai-PRL-105-2010}.

\textbf{5. Acknowledgments}

We thank the Department of Science and Technology, India for financial
support.

\end{document}